\documentclass[12pt,a4paper]{article}
 
\usepackage[T1]{fontenc}
\usepackage[utf8]{inputenc}
\usepackage{graphicx}
\usepackage{float}
\usepackage[margin=2.5cm]{geometry}
\usepackage{microtype}
\usepackage{mathptmx}
\usepackage{setspace}
\usepackage{amsmath,amssymb}
\usepackage{booktabs}
\usepackage{array}
\usepackage{tabularx}
\usepackage{caption}
\usepackage{authblk}
\usepackage{abstract}
\usepackage{titlesec}
\usepackage{lineno}
\usepackage{xcolor}
 
\definecolor{MidnightBlue}{HTML}{000066} 
\definecolor{DarkOrange}{HTML}{CC5500}
\definecolor{Rust}{HTML}{B7410E} 
\definecolor{DeepRust}{HTML}{8B310A}
\definecolor{RustOrange}{HTML}{662200}
 
\usepackage[
    colorlinks=true,
    citecolor=RustOrange,  
    linkcolor=RustOrange,
    urlcolor=teal,
    pdftitle={Long-Range Correlation in Code Commit Dynamics},
    pdfauthor={Goran Mitevski}
]{hyperref}
 
\usepackage[authoryear,round]{natbib}
\usepackage{parskip}
 
\makeatletter
\renewcommand\NAT@open{\textcolor{RustOrange}{(}} 
\renewcommand\NAT@close{\textcolor{RustOrange}{)}} 
\makeatother
 
\makeatletter
\def\NAT@nmfmt#1{#1} 
\makeatother
 
\titleformat{\section}{\large\bfseries\scshape}{\thesection.}{0.6em}{}[\vspace{-2pt}\rule{\linewidth}{0.4pt}]
\titleformat{\subsection}{\normalsize\bfseries}{\thesubsection}{0.6em}{}
\titleformat{\subsubsection}{\normalsize\bfseries\itshape}{\thesubsubsection}{0.6em}{}
\titlespacing{\section}{0pt}{18pt}{6pt}
\titlespacing{\subsection}{0pt}{12pt}{4pt}
\titlespacing{\subsubsection}{0pt}{10pt}{2pt}
 
\numberwithin{equation}{section}
 
\newcolumntype{L}[1]{>{\raggedright\arraybackslash}p{#1}}
\newcolumntype{C}[1]{>{\centering\arraybackslash}p{#1}}
 
\begin{document}

\title{\textbf{Long-Range Correlation in Code Commit Dynamics\\
as a Novel Indicator of Software Product Stability:\\
A Detrended Fluctuation Analysis Study}\\[6pt]
\normalsize\textit{Fractal structure as a code-inspection-free process quality metric}}

\author[1]{\textbf{Goran Mitevski}}

\date{January, 2023}
\maketitle
\thispagestyle{empty}

\begin{abstract}
This work proposes the fractal scaling exponent $\alpha$, estimated via Detrended
Fluctuation Analysis (\textsc{DFA}) on the unaggregated time series of lines
of code added per commit event in a software repository, as a novel process-level indicator of software
product stability. The proposal rests on the hypothesis that stable software
products arise from development processes characterised by long-range
temporal correlations in commit behaviour: each code addition is shaped not
only by the immediately preceding commits but by patterns extending weeks or
months into the past and anticipating work to be done in the future. This hypothesis is tested on two non-overlapping 712-day time series of lines of code added per commit event, drawn from a closed-source software organisation and labeled as stable and unstable by the lead engineer on the basis of
crash-analytics data. Applied to these series, \textsc{DFA} yields
$\alpha = 0.70$ ($n_{\min}=16$) for the stable period and $\alpha = 0.57$
for the unstable period, with all estimates substantially above the
shuffled-surrogate baseline ($\alpha \approx 0.50 \pm 0.01$). Results are
robust to three parameterisations ($n_{\min} \in \{4, 16, 48\}$) and
validated against 1,000 surrogate time series per condition. Remarkably, the
unstable period generated 3.2 times more commit events than the stable period, yet
exhibited lower long-range memory, demonstrating that commit volume alone
does not predict stability, and that the temporal \emph{organisation} of
development activity is the key variable. This result can be situated in the
broader literature on fractality in human creative production, discuss
methodological limitations, and outline a research programme for deploying
$\alpha$ as a continuous code-health indicator in version-control pipelines.

\end{abstract}

\vspace{1em}
\hrule
\vspace{1.5em}

\section{Introduction}

Software quality is expensive  to measure. The dominant methods of static code analysis, testing, and code review require access to source code or the execution of test suites. Alternatively, production metrics like crash analytics require the prior release of potentially defective software. A metric derivable from the temporal pattern of development activity alone (from the stream of
timestamped commits to a repository) would have considerable practical
value: it would require no code review, no test execution, no natural
language processing, and could be monitored continuously from any
version-control log.

This work proposes such a metric: the fractal scaling exponent $\alpha$ on the unaggregated time series of lines of code added per commit event in a software repository, estimated via Detrended Fluctuation Analysis (\textsc{DFA}). The core hypothesis is that the temporal organisation of
commit behaviour encodes information about the quality of the software being
produced. Specifically, that stable, high-quality software is
the product of development processes characterised by long-range temporal
correlations: each commit is shaped not only by the immediately preceding
commits but by patterns extending weeks or months into the past, and in some
anticipatory sense by work expected in the future. This interdependence
across time scales is precisely what \textsc{DFA}'s scaling exponent
$\alpha$ quantifies.

The argument has an intuitive foundation. Software architecture is
inherently hierarchical and interdependent: a change to a core module today
should be consistent with design decisions made three months ago and should
anticipate module extensions planned for next month. A developer with deep
involvement in a codebase will make commits that reflect this multi-scale
awareness, commits that are, in a precise mathematical sense, correlated
with the distant past and future of the time series. A developer without
such involvement, or a team under deadline pressure patching problems
reactively, will make commits that are more locally determined, producing a
time series with weaker long-range structure.

\textsc{DFA}, introduced by \citet{peng1994} for the analysis of DNA
nucleotide sequences, is designed precisely to detect such long-range
structure in the presence of non-stationarities that would confound simpler
autocorrelation measures. Its output, the scaling exponent $\alpha$, is a
single number: $\alpha \approx 0.5$ indicates white noise (no memory);
$\alpha \in (0.5, 1.0]$ indicates positive long-range correlations of
increasing strength; $\alpha > 1.0$ indicates non-stationary behaviour. This work
proposes that the region $\alpha \in (0.5, 1.0]$---and specifically the
distance of $\alpha$ above 0.5---is a continuous indicator of commit-process
coherence and, by extension, of likely software product stability.

\subsection{Scope and Contribution}

This paper makes four contributions. First, it proposes $\alpha$ from
\textsc{DFA} on commit time series as a novel, code-inspection-free indicator
of software process quality. Second, it provides an empirical test of this
proposal against externally validated stability labels from a real
organisation. Third, it addresses the methodological challenges of applying
\textsc{DFA} to commit data, including non-stationarity, length effects, and
window-size sensitivity, with robust parameterisation and surrogate
validation. Fourth, it situates the finding within the broader literature on
fractality in human creative production and opens a research programme
connecting complexity science with software engineering practice.

\section{Background and Related Work}

\subsection{Fractality and Long-Range Correlation in Human Creative Production}

The concept of statistical self-similarity (fractal structure) has been
applied to human language, cognition, and action across several decades.
\citet{mandelbrot1983} observed that word-frequency distributions in
natural language follow power laws. \citet{gilden1995} found that simple
reaction times exhibit $1/f$ noise---the power spectrum associated with
long-range correlations and $\alpha \approx 1$ in \textsc{DFA} terms. The
broader claim \citep{vanorden2003,riley2002} is that long-range correlations
in performance time series are a signature of flexible, adaptive engagement
with a task: a system that is too rigid produces trivially correlated
outputs, while one that is too random produces white noise. The productive
middle ground is $1/f$ noise.

In the domain of language and narrative, several studies are directly
relevant. \citet{butner2008} analysed the fractal structure of
conversational stories and found that more fractal stories were rated as
higher quality by listeners---a striking parallel to the findings of this work that more
fractal commit processes produce more stable software.
\citet{drozdz2016} examined long-range correlations in famous literary
texts, finding that canonical works exhibit stronger multifractal structure
than lesser-known texts. \citet{bhan2006} found similar structure in Korean
literary corpora. These findings collectively suggest that long-range temporal
organisation is a hallmark of high-quality human creative production across
modalities---and that software development, as a structured creative activity,
may be no exception.

To the best of the author's knowledge, no previous study has applied \textsc{DFA} to commit-level code-addition time series as a software quality indicator.

\subsection{Detrended Fluctuation Analysis: Origins and Applications}

\textsc{DFA} was introduced by \citet{peng1994} to study the long-range
correlation properties of DNA nucleotide sequences. Its key advantage over
earlier methods (Hurst exponent, spectral analysis) is robustness to
non-stationarity: embedded polynomial trends are removed locally within each
window, preventing spurious long-range correlation estimates that would arise
from global trends. \citet{bashan2008} provide a systematic comparison of
\textsc{DFA} and related methods, confirming its robustness across a wide
range of conditions. \citet{goldberger2000} established the normative
framework for interpreting $\alpha$ values in the physiological context,
which this work adapts for software.

\textsc{DFA} has been applied in physiology (cardiac interbeat intervals,
gait, EEG), financial time series, environmental science, and
cognitive science (reaction time series, attention dynamics). Its track
record across these domains provides a methodological foundation for the
present application.

\subsection{Creativity, Stability, and the Software Development Process}

\citet{boden2007} defines creativity as the production of ideas or
artefacts that are new, surprising, and valuable. In the software context,
value is reflected in product stability and user satisfaction.
\citet{yu2020} provide an empirical foundation for using stability as a
value proxy: stable software products (lower crash rates, faster response
times) receive higher user ratings. This establishes the chain: long-range
commit correlations $\to$ stable software $\to$ higher user value $\to$
greater product success.

\subsection{The Scaling Exponent Interpretation Framework}

The \textsc{DFA} scaling exponent $\alpha$ admits the following interpretive
framework \citep{goldberger2000,bashan2008}:

\begin{center}
\small
\begin{tabular}{lp{8cm}}
\toprule
$\alpha \approx 0.5$ & Uncorrelated white noise (memoryless process) \\
$\alpha < 0.5$ & Anti-correlated (negatively persistent) \\
$0.5 < \alpha < 1.0$ & Positively long-range correlated (persistent) \\
$\alpha \approx 1.0$ & Pink ($1/f$) noise; signature of complex adaptive systems \\
$\alpha > 1.0$ & Non-stationary; Brownian-like or unbounded \\
$\alpha \approx 1.5$ & Fractional Brownian motion \\
\bottomrule
\end{tabular}
\end{center}

\noindent For commit time series, this work expects $\alpha \in (0.5, 1.0]$ for
any sustained development process. This work hypothesis is that the distance of
$\alpha$ above 0.5 co-varies with product stability.

\section{Data}

\subsection{Source and Labeling}

Commit-level time series were extracted from the complete GitHub activity
log of a small, globally distributed closed-source software (\textsc{css})
organisation with more than five years of continuous operation. Two
non-overlapping periods of 712 days each were identified and labeled by the
lead software engineer as \textit{stable} and \textit{unstable}. The labeling
criterion was the frequency of software crashes (instances in which the
software became unresponsive) as reported by Google Analytics and equivalent
services. This is an operationally clear criterion grounded in user-facing
behaviour rather than developer judgment of code quality.

\subsection{Time Series Construction}

The time series were constructed directly from the raw commit log: each commit
event constitutes one observation, ordered chronologically by its UTC-aware
timestamp. The value at each observation is the number of lines of code added
in that commit. Key descriptive statistics: the stable period contained 976 commit events across 712 days (mean: 1.37 commits/day); the unstable period contained 3,129 commit events
across 712 days (mean: 4.39 commits/day)---approximately 3.2 times more
commit activity. This asymmetry is a central feature of the comparison.

\subsection{Crossover Analysis}

Following the methodological framework of \citet{peng1995}, both time series were inspected for crossovers, which are pronounced changes in scaling behavior across different timescales. In \textsc{DFA}, a crossover in the $\log F(n)$ vs.\ $\log n$ plot typically indicates a transition between distinct scaling regimes, potentially signaling a shift in the underlying system dynamics \citep{ogata2006} or the influence of non-stationary trends that could confound the interpretation of a single scaling exponent \citep{hu2001}. 

Visual and linear regression analysis of the fluctuation functions revealed that while minor local fluctuations are present (typical of empirical human-activity time series) the scaling behavior remains sufficiently consistent across the examined window ranges. The absence of systematic breaks or slope shifts suggests that the code-commit dynamics are governed by a dominant monofractal process rather than a multifractal one. Therefore, the application of a single scaling exponent $\alpha$ is a parsimonious and appropriate descriptor for characterizing both periods (see \hyperref[app:crossover]{Appendix~\ref*{app:crossover}} for the detailed visual diagnostic).

\section{Methods}

\subsection{Detrended Fluctuation Analysis}

Let $x(i)$, $i = 1, \ldots, N$ be the commit-level time series.
\textsc{DFA} proceeds in three stages.

\subsubsection{Stage 1: Integration}

The series is mean-centred and integrated:
\begin{equation}
  y(k) = \sum_{i=1}^{k} \bigl[x(i) - \langle x \rangle\bigr]
  \label{eq:integrate}
\end{equation}
where $\langle x \rangle$ is the series mean.

\subsubsection{Stage 2: Local Detrending}

The integrated series $y$ is divided into non-overlapping windows of length
$n$. Within each window, a least-squares polynomial of degree $d=1$ is
fitted to estimate the local trend $y_n(k)$. The root-mean-square
fluctuation is:
\begin{equation}
  F(n) = \sqrt{\frac{1}{N} \sum_{k=1}^{N} \bigl(y(k) - y_n(k)\bigr)^2}
  \label{eq:fluctuation}
\end{equation}
Local detrending suppresses non-stationarities that would otherwise inflate
$\alpha$ estimates---the key innovation of \textsc{DFA} relative to earlier
fluctuation methods.

\subsubsection{Stage 3: Scaling Analysis}

Stage 2 is repeated over a range of window sizes $n$. If $F(n)$ scales as a
power law,
\begin{equation}
  F(n) \propto n^{\alpha}
  \label{eq:scaling}
\end{equation}
then $\alpha$ is estimated as the slope of the $\log F(n)$ vs.\ $\log n$
regression. Linearity of the log-log plot is a necessary condition for
fractal scaling and was verified visually and quantitatively for both series.

\subsection{Parameterisation Strategy}

To assess robustness, three parameterisations were tested:

\textbf{Parameterisation~1} (recommended): $n_{\min}=16$,
$n_{\max}=N/4$. For the stable series ($N=976$ events), $n_{\max}=244$,
yielding 13 distinct window sizes. For the unstable series ($N=3{,}129$),
$n_{\max}=782$, yielding 18 window sizes. The $n_{\min}=16$ choice
corresponds to approximately two weeks of stable-period activity.

\textbf{Parameterisation~2}: $n_{\min}=48$, $n_{\max}=N/4$. A larger
minimum window reduces sensitivity to short-timescale fluctuations.

\textbf{Parameterisation~3} (exploratory): $n_{\min}=4$,
$n_{\max}=N/2$. An aggressive parameterisation serving as a stress test.

\subsection{Surrogate Validation}

To confirm that observed $\alpha$ values reflect genuine long-range
correlations rather than statistical artefacts, 1,000 surrogate series were
generated for each condition by randomly shuffling the commit-level
values without replacement \citep{theiler1992,moulder2018}. Shuffling
destroys temporal order---and thus all serial correlation---while preserving
the marginal distribution. \textsc{DFA} was applied to each surrogate series.
The resulting distribution provides an empirical null distribution.

An observed $\alpha$ is considered to reflect genuine long-range memory if
and only if it lies outside the 95\% confidence interval of the surrogate
distribution. Surrogate series consistently yield $\alpha \approx 0.50 \pm
0.01$.

\subsection{Evaluation Criteria}

Long-range memory is inferred when: (1) $\alpha > 0.5$ in the original
series; (2) $\alpha < 1.0$; (3) $\alpha$ lies outside the surrogate 95\%
CI; and (4) the surrogate mean is $\approx 0.5$. Criterion (3) is
decisive. Comparison between periods is made directly on $\alpha$ values;
a larger $\alpha$ in the stable period, consistent across all
parameterisations, supports the hypothesis.

\section{Results}

\subsection{Stable Period}

Table~\ref{tab:DFA} reports $\alpha$ estimates for the stable period across
all parameterisations. Under the recommended parameterisation
($n_{\min}=16$), $\alpha = 0.70$. Under $n_{\min}=48$, $\alpha = 0.66$.
Under the exploratory $n_{\min}=4$, $\alpha = 0.67$. All three estimates
lie within the long-range memory range $(0.5, 1.0)$ and all lie well outside
the corresponding 95\% CI of the surrogate distribution.

The consistency across parameterisations (range: $0.66$--$0.70$) indicates
that the long-range memory in the stable period is robustly present across
timescales from days to months. Values in the range $0.66$--$0.70$ correspond
to moderate persistent correlations---below the $1/f$ boundary ($\alpha=1.0$)
but substantially above white noise.

\subsection{Unstable Period}

Table~\ref{tab:DFA} reports $\alpha$ estimates for the unstable period across
all parameterisations. For the unstable period, $\alpha = 0.57$ ($n_{\min}=16$), $\alpha = 0.62$
($n_{\min}=48$), and $\alpha = 0.56$ ($n_{\min}=4$). All three estimates
exceed the surrogate upper bound, indicating long-range memory above chance.
However, all three are lower than the corresponding stable-period estimates
by a consistent margin of $0.04$--$0.14$ units. The unstable period values
cluster closer to 0.5, indicating weaker (though still present) long-range
memory.

\begin{table}[ht]
\centering
\caption{\textsc{DFA} scaling exponents with surrogate validation across
  three parameterisations. Surrogate values are mean $\pm$ 95\% CI across
  1,000 shuffled realisations. $\dagger$ Non-recommended exploratory
  parameterisation.}
\label{tab:DFA}
\small
\begin{tabular}{L{4.2cm} C{1.3cm} C{2.0cm} C{1.3cm} C{2.0cm} C{1.3cm}}
\toprule
\textbf{Parameterisation}
  & \textbf{Stable $\alpha$}
  & \textbf{Stable\newline surrogates}
  & \textbf{Unstable $\alpha$}
  & \textbf{Unstable\newline surrogates}
  & \textbf{$\Delta\alpha$} \\
\midrule
$n_{\min}=16$, $n_{\max}=N/4$
  & 0.70 & $[0.49,\,0.50]$ & 0.57 & $[0.49,\,0.50]$ & $+0.13$ \\
$n_{\min}=48$, $n_{\max}=N/4$
  & 0.66 & $[0.49,\,0.50]$ & 0.62 & $[0.49,\,0.50]$ & $+0.04$ \\
$n_{\min}=4$, $n_{\max}=N/2$ $\dagger$
  & 0.67 & $[0.50,\,0.51]$ & 0.56 & $[0.50,\,0.50]$ & $+0.11$ \\
\bottomrule
\end{tabular}
\end{table}

\subsection{Commit frequency volume vs.\ Long-Range Memory}

The most striking finding is the dissociation between commit frequency volume and
$\alpha$. The unstable period generated 3,129 commits versus 976 in the
stable period (3.2 times more activity) yet consistently produced lower
$\alpha$. This directly refutes the hypothesis that more development activity
produces more stable software. The temporal organisation of that activity,
not its volume, is the discriminating variable.

This pattern is consistent with the `thrashing' phenomenon well known in
software project management: teams under deadline pressure commit frequently
but without deep integration, producing many small reactive fixes that
collectively degrade rather than improve software quality. The \textsc{DFA}
exponent may be capturing exactly this difference: architecturally integrated
development (high $\alpha$) versus reactive, patch-driven development (low
$\alpha$).

\subsection{Comparison to Adjacent Literature}

The stable period $\alpha$ values ($0.66$--$0.70$) are comparable to those
found in high-quality literary texts \citep{drozdz2016}, conversational
stories rated as high quality \citep{butner2008}, and physiological signals
from healthy, adaptive systems \citep{goldberger2000}. The unstable period
values ($0.56$--$0.62$) are closer to values found in more repetitive or
less structurally complex materials. This convergence across domains
strengthens the interpretation of $\alpha$ as a general indicator of
structured, adaptive creative production.

\section{Discussion}

\subsection{Interpretation of the Core Finding}

The central finding supports the proposed framework: stable software
development is characterised by commit patterns in which current decisions
are shaped by a memory stretching weeks and months into the past. This
coherence manifests as statistical self-similarity in the commit time series.
In practical terms, the rhythm of development is not reset at arbitrary
boundaries: a productive week tends to follow a productive month, and a slow
day tends to follow a slow week, with these tendencies persisting across the
full range of timescales examined.

The fact that both periods exceed the surrogate baseline (both $\alpha > 0.5
+ \text{surrogate bound}$) indicates that commit dynamics are not random in
either condition---some degree of multi-scale structure is present in all
sustained development processes. What distinguishes stable from unstable
development is the strength of this structure.

\subsection{Frequency Volume vs.\ Quality: A Fundamental Asymmetry}

The 3.2$\times$ greater commit frequency volume of the unstable period refutes the
trivial hypothesis that more activity produces better software. A process
characterised by reactive, patch-driven commits will produce high volume but
weak long-range structure. Each commit is determined primarily by the
immediately preceding problem, not by the multi-week and multi-month patterns
of the development trajectory.

This interpretation connects to the established concept of technical debt in
software engineering: short-term expedient decisions that solve an immediate
problem without considering long-term consequences. A low $\alpha$ may be an
early indicator of technical debt accumulation, identifiable from commit
timestamps alone, before its consequences become visible in crash statistics
or user complaints.

\subsection{Methodological Considerations}

Several methodological issues deserve explicit discussion. First, the choice
of minimum window size in \textsc{DFA} is consequential: the $n_{\min}=48$
result for the unstable period ($\alpha=0.62$) is higher than other
parameterisations, suggesting that at longer timescales, the unstable period
shows more structure than at shorter timescales. This could indicate that
while short-term commit patterns are reactive, there is some longer-term
periodicity at the multi-month scale.

Second, the series differ in length (in events, not days). This work adjusted
minimum window sizes proportionally ($n_{\min}=48 \approx 3 \times 16$) to
account for this difference, and found that the stable-period $\alpha$
remains higher even under this adjustment.

Third, \textsc{DFA} as applied here assumes monofractal scaling. This
assumption was verified through crossover analysis. If future data exhibit
crossovers, multifractal \textsc{DFA} (\textsc{MFDFA}) would be required,
yielding a spectrum of exponents rather than a single $\alpha$.

Fourth, the \textsc{DFA} analysis is based solely on lines of code \textit{added} per
commit; deletions were not included. To the best of the author's knowledge,
no prior study has examined the fractal structure of code-deletion dynamics,
and additions alone capture the primary signal of constructive development
activity required to test the core hypothesis advanced here.

Fifth, the stability label was assigned by a single domain expert, grounded
in objective crash analytics. Future work should use multiple independent
raters and finer-grained stability measures to enable continuous rather than
binary analysis.

\subsection{Practical Deployment: $\alpha$ as a Continuous Code-Health Metric}

The practical appeal of $\alpha$ lies in its computational simplicity and
its inputs. Computing $\alpha$ requires only a time series of commit
timestamps, data that any organisation using version control generates as a
by-product of normal development activity. The \textsc{DFA} algorithm is
$\mathcal{O}(N \log N)$ and takes seconds for time series of the lengths
considered here. It requires no code access, no build system, no test suite,
and no language-specific tooling.

A practical monitoring system could compute $\alpha$ on a rolling 30-, 60-,
or 90-day window of commit data, updating daily. A decline in $\alpha$, particularly a sustained decline across multiple window sizes, would serve
as an early-warning signal that the development process is losing its
multi-scale coherence. For integration into \textsc{CI/CD} pipelines,
$\alpha$ computation could be triggered on each push to the main branch,
with the result logged alongside conventional quality metrics.

\subsection{Connections to Creativity Research and Team Science}

This work is positioned within \citet{batey2012}'s taxonomic framework as
a team-level, process-facet, objective-measurement approach. It measures the
temporal organisation of the development process, the fractal texture of
collective creative production, rather than the creativity of individual
developers or the novelty of specific features.

In this sense, $\alpha$ is a measure of creative \textit{coherence} rather
than creative \textit{output}. A high $\alpha$ does not guarantee that good
software will result, nor does a low $\alpha$ preclude good outcomes. But on
average, across the timescales of sustained software development projects, this work
proposes that higher $\alpha$ is associated with more stable outcomes.

\subsection{Future Directions}

Multifractal \textsc{DFA} (\textsc{MFDFA}) and Wavelet Transform Modulus
Maxima (\textsc{WTMM}) analysis would reveal whether the full spectrum of
scaling exponents, not only the dominant $\alpha$, varies between conditions.
Cross-team complexity matching, examining whether teams on the same project
have commit time series with different $\alpha$ values, could index
coordination difficulties, analogous to complexity matching in motor
coordination research \citep{marmelat2012}. 

An important avenue of research is the difference in long-term memory and complexity between teams that produce software solely through human effort and teams that rely heavily on generative LLM-based AI for code generation, and whether higher long-range correlation in lines of code added per commit remains a valid indicator of software stability when the commit process is driven predominantly by LLM-based generation.
 
Combining \textsc{DFA} on commits with the Lyapunov exponent analysis of
discussion dynamics, opens the possibility of a two-dimensional characterisation of the creative state of a software development team: the fractal coherence of the code process and the
chaotic vitality of the surrounding discussion. Together, these metrics
can provide a richer picture than either alone.

\section{Conclusion}

This work proposes the \textsc{DFA} scaling exponent $\alpha$ of the unaggregated time series of lines of code added per commit event as a novel, code-inspection-free indicator of
software product stability, and provides an empirical test of this
proposal using data from a real distributed software organisation. The stable
software period yields $\alpha = 0.66$--$0.70$ across parameterisations; the
unstable period yields $\alpha = 0.56$--$0.62$; both are validated against
1,000 surrogate series per condition. The stable period's higher $\alpha$
persists despite having only 31\% as many commit events as the unstable
period, suggesting that the temporal organisation of development activity
may be a more informative signal than commit frequency volume alone.

These findings add software commit dynamics to the class of human creative
productions whose fractal structure co-varies with quality, alongside
literary texts, conversational stories, and physiological signals of adaptive
system function. \textsc{DFA} on commit time series is computationally
trivial, platform-agnostic, and producible from any version-control log. This work
proposes it as a complementary metric to existing code-quality tools,
deployable continuously within any development organisation, and as a
foundation for a new research programme connecting complexity science with
software engineering.

\newpage
\section*{Data Availability}

The commit-level time series analysed in this study are made available as
two comma-separated-values (CSV) files deposited on Zenodo under a Creative
Commons Attribution 4.0 International (CC~BY~4.0) licence
\citep{mitevski2026data}:

\begin{itemize}
    \item \texttt{unstable\_period\_code\_additions.csv} — 3,129 commit events
          recorded during the 712-day unstable period (mean: 4.39
          commits/day). Each row contains a UTC-aware ISO~8601 timestamp
          (\texttt{time\_point}) and the number of lines of code added in that
          commit event (\texttt{additions}).

    \item \texttt{stable\_period\_code\_additions.csv} — 976 commit events
          recorded during the 712-day stable period (mean: 1.37
          commits/day). Columns are identical to those of the unstable-period
          file.
\end{itemize}

The source repository and organisational identity of the software company
remain confidential to protect commercially sensitive information; the
commit-level time series provided are sufficient to reproduce all
statistical analyses reported in this study. The dataset is available at
\url{https://doi.org/10.5281/zenodo.19986248}.


\section*{Declaration of generative AI and AI-assisted technologies}

This work was created by the author in January 2023. 
In May 2026, LLM based generative AI was used to improve the readability and language of the original work.
The author takes full responsibility for the content of this publication.

\section*{Acknowledgement}

First and foremost, I would like to thank my wife, Marija Efremova, whose incredible talent and work were the inspiration and driving force behind this work. I would also like to acknowledge Dr. Travis Wiltshire for his exceptional support, knowledge, and guidance in Complex Systems, as well as for mentoring my thesis work, of which this work was a part.
I would like to acknowledge my mother Ana Mitevska, a talented mathematician and a Germanist, Velin Mitevski, whose work in expert and decision support systems in the 1970s and 1990s sparked my interest in this problem space. Last but not least, I would like to acknowledge Mitko Efremov, whose work and involvement in centralised automation design, and engineering sustainable mass water supply systems were an inspiration behind some of the insights in this work.

\bibliographystyle{apalike}

\begin{thebibliography}{}

\bibitem[Barabási(2005)]{barabasi2005}
Barabási, A.-L. (2005).
\newblock The origin of bursts and heavy tails in human dynamics.
\newblock \textit{Nature}, \textbf{435}(7039).
\newblock \url{https://doi.org/10.1038/nature03459}

\bibitem[Bashan et al.(2008)]{bashan2008}
Bashan, A., Bartsch, R., Kantelhardt, J.\,W., \& Havlin, S. (2008).
\newblock Comparison of detrending methods for fluctuation analysis.
\newblock \textit{Physica A}, \textbf{387}(21), 5080--5090.
\newblock \url{https://doi.org/10.1016/j.physa.2008.04.023}

\bibitem[Batey(2012)]{batey2012}
Batey, M. (2012).
\newblock The measurement of creativity: From definitional consensus to the introduction of a new heuristic framework.
\newblock \textit{Creativity Research Journal}, \textbf{24}(1), 55--65.
\newblock \url{https://doi.org/10.1080/10400419.2012.649181}

\bibitem[Bhan et al.(2006)]{bhan2006}
Bhan, J., Kim, S., Kim, J., Kwon, Y., Yang, S., \& Lee, K. (2006).
\newblock Long-range correlations in Korean literary corpora.
\newblock \textit{Chaos, Solitons \& Fractals}, \textbf{29}(1), 69--81.
\newblock \url{https://doi.org/10.1016/j.chaos.2005.08.214}

\bibitem[Boden(2007)]{boden2007}
Boden, M.\,A. (2007).
\newblock Creativity in a nutshell.
\newblock \textit{Think}, \textbf{5}(15), 83--96.
\newblock \url{https://doi.org/10.1017/S147717560000230X}

\bibitem[Butner et al.(2008)]{butner2008}
Butner, J., Pasupathi, M., \& Vallejos, V. (2008).
\newblock When the facts just don't add up: The fractal nature of conversational stories.
\newblock \textit{Social Cognition}, \textbf{26}, 670--699.
\newblock \url{https://doi.org/10.1521/soco.2008.26.6.670}

\bibitem[Couger \& Dengate(1996)]{couger1996}
Couger, J.~D., \& Dengate, G. (1996).
\newblock Measurement of creativity of IS products.
\newblock \textit{Creativity and Innovation Management}, \textit{5}(4), 262--272.
\newblock \url{https://doi.org/10.1111/j.1467-8691.1996.tb00152.x}

\bibitem[Drożdż et al.(2016)]{drozdz2016}
Drożdż, S., Oświęcimka, P., Kulig, A., Kwapień, J., Bazarnik, K., Grabska-Gradzińska, I., Rybicki, J., \& Stanuszek, M. (2016).
\newblock Quantifying origin and character of long-range correlations in narrative texts.
\newblock \textit{Information Sciences}, \textbf{331}, 32--44.
\newblock \url{https://doi.org/10.1016/j.ins.2015.10.023}

\bibitem[Feldman(2012)]{feldman2012}
Feldman, D.\,P. (2012).
\newblock \textit{Chaos and Fractals: An Elementary Introduction}.
\newblock Oxford University Press.

\bibitem[Gilden et al.(1995)]{gilden1995}
Gilden, D.\,L., Thornton, T., \& Mallon, M.\,W. (1995).
\newblock $1/f$ noise in human cognition.
\newblock \textit{Science}, \textbf{267}(5205), 1837--1839.
\newblock \url{https://doi.org/10.1126/science.7892611}

\bibitem[Goldberger et al.(2000)]{goldberger2000}
Goldberger, A.\,L., Amaral, L.\,A.\,N., Glass, L., Hausdorff, J.\,M., Ivanov, P.\,C., Mark, R.\,G., Mietus, J.\,E., Moody, G.\,B., Peng, C.-K., \& Stanley, H.\,E. (2000).
\newblock PhysioBank, PhysioToolkit, and PhysioNet.
\newblock \textit{Circulation}, \textbf{101}(23), e215--e220.
\newblock \url{https://doi.org/10.1161/01.CIR.101.23.e215}

\bibitem[Guastello(1998)]{guastello1998}
Guastello, S.\,J. (1998).
\newblock Creative problem solving groups at the edge of chaos.
\newblock \textit{The Journal of Creative Behavior}, \textbf{32}(1), 38--57.
\newblock \url{https://doi.org/10.1002/j.2162-6057.1998.tb00805.x}

\bibitem[Hu et al.(2001)]{hu2001}
Hu, K., Ivanov, P.~C., Chen, Z., Carpena, P., and Stanley, H.~E. (2001).
\newblock Effect of trends on detrended fluctuation analysis.
\newblock \textit{Physical Review E}, \textbf{64}(1), 011114.
\newblock \url{https://doi.org/10.1103/PhysRevE.64.011114}

\bibitem[Mandelbrot(1983)]{mandelbrot1983}
Mandelbrot, B.\,B. (1983).
\newblock \textit{The Fractal Geometry of Nature}.
\newblock Freeman.

\bibitem[Marmelat \& Delignières(2012)]{marmelat2012}
Marmelat, V., \& Delignières, D. (2012).
\newblock Strong anticipation: Complexity matching in interpersonal coordination.
\newblock \textit{Experimental Brain Research}, \textbf{222}, 137--148.
\newblock \url{https://doi.org/10.1007/s00221-012-3202-9}

\bibitem[Mitevski \& Efremova(2026)]{mitevski2026data}
Mitevski, G., \& Efremova, M. (2026).
\newblock Commit-level time series for stable and unstable software periods
  [Data set].
\newblock \textit{Zenodo}.
\newblock \url{https://doi.org/10.5281/zenodo.19986248}

\bibitem[Moulder et al.(2018)]{moulder2018}
Moulder, R.\,G., Boker, S.\,M., Ramseyer, F., \& Tschacher, W. (2018).
\newblock Determining synchrony between behavioral time series.
\newblock \textit{Psychological Methods}, \textbf{23}, 757--773.
\newblock \url{https://doi.org/10.1037/met0000172}

\bibitem[Nelson et al.(2010)]{nelson2010}
Nelson, C., Brummel, B., Grove, D.\,F., Jorgenson, N., Sen, S., \& Gamble, R.\,C. (2010).
\newblock Measuring creativity in software development.
\newblock \textit{Proceedings of ICCC-10}, 205--214.

\bibitem[Ogata et al.(2006)]{ogata2006}
Ogata, H., Tokuyama, K., Nagasaka, S., Ando, A., Kusaka, I., Sato, A., ... and Ishibashi, S. (2006).
\newblock Long-range negative correlation of glucose dynamics in humans and its breakdown in diabetes mellitus.
\newblock \textit{American Journal of Physiology-Regulatory, Integrative and Comparative Physiology}, \textbf{291}(6), R1638--R1643.
\newblock \url{https://doi.org/10.1152/ajpregu.00241.2006}

\bibitem[Paulson et al.(2004)]{paulson2004}
Paulson, J., Succi, G., \& Eberlein, A. (2004).
\newblock An empirical study of open-source and closed-source software products.
\newblock \textit{IEEE Transactions on Software Engineering}, \textbf{30}, 246--256.
\newblock \url{https://doi.org/10.1109/TSE.2004.1274044}

\bibitem[Peng et al.(1994)]{peng1994}
Peng, C.-K., Buldyrev, S.\,V., Havlin, S., Simons, M., Stanley, H.\,E., \& Goldberger, A.\,L. (1994).
\newblock Mosaic organization of DNA nucleotides.
\newblock \textit{Physical Review E}, \textbf{49}(2), 1685--1689.
\newblock \url{https://doi.org/10.1103/PhysRevE.49.1685}

\bibitem[Peng et al.(1995)]{peng1995}
Peng, C.-K., Havlin, S., Stanley, H.~E., and Goldberger, A.~L. (1995).
\newblock Quantification of scaling exponents and crossover phenomena in nonstationary heartbeat time series.
\newblock \textit{Chaos}, \textbf{5}(1), 82--87.
\newblock \url{https://doi.org/10.1063/1.166141}

\bibitem[Riley \& Turvey(2002)]{riley2002}
Riley, M.\,A., \& Turvey, M.\,T. (2002).
\newblock Variability and determinism in motor behavior.
\newblock \textit{Journal of Motor Behavior}, \textbf{34}(2), 99--125.
\newblock \url{https://doi.org/10.1080/00222890209601934}

\bibitem[Theiler et al.(1992)]{theiler1992}
Theiler, J., Eubank, S., Longtin, A., Galdrikian, B., \& Farmer, J.\,D. (1992).
\newblock Testing for nonlinearity in time series: The method of surrogate data.
\newblock \textit{Physica D}, \textbf{58}(1), 77--94.
\newblock \url{https://doi.org/10.1016/0167-2789(92)90102-S}

\bibitem[Van~Orden et al.(2003)]{vanorden2003}
Van~Orden, G.\,C., Holden, J.\,G., \& Turvey, M.\,T. (2003).
\newblock Self-organization of cognitive performance.
\newblock \textit{Journal of Experimental Psychology: General}, \textbf{132}(3), 331--350.
\newblock \url{https://doi.org/10.1037/0096-3445.132.3.331}

\bibitem[Varela et al.(2016)]{varela2016}
Varela, M., Vigil, L., Rodriguez, C., Vargas, B., and García-Carretero, R. (2016).
\newblock Delay in the detrended fluctuation analysis crossover point as a risk factor for type 2 diabetes mellitus.
\newblock \textit{Journal of Diabetes Research}, \textbf{2016}, Article ID 9361958.
\newblock \url{http://dx.doi.org/10.1155/2016/9361958}

\bibitem[Yu et al.(2020)]{yu2020}
Yu, M., Zhou, R., Cai, Z., Tan, C.-W., \& Wang, H. (2020).
\newblock Unravelling the relationship between response time and user experience in mobile applications.
\newblock \textit{Internet Research}, \textbf{30}(5), 1353--1382.
\newblock \url{https://doi.org/10.1108/INTR-05-2019-0223}

\end{thebibliography}


\newpage
\appendix
\renewcommand{\thesection}{\Alph{section}}
\setcounter{section}{8}   

\section*{Appendices}
\addcontentsline{toc}{section}{Appendices}

\section{Crossover Check}
\label{app:crossover}

This appendix details the diagnostic procedure used to verify the absence of significant crossovers in the \textsc{DFA} log-log plots for the \textsc{css} `stable' and `unstable' commit-level time series.

\subsection*{The Significance of Crossovers}

In the context of \textsc{DFA}, a crossover is defined as a change in the scaling exponent $\alpha$ at a specific characteristic scale $n$. According to \citet{peng1995}, such phenomena suggest that the system's underlying dynamics change as a function of the observation window. Furthermore, \citet{hu2001} demonstrated that crossovers can emerge as artifacts of non-stationarity (e.g., linear or power-law trends) in the data. 

While empirical data rarely yields a perfectly linear scaling law, identifying a sufficiently consistent linear region is a prerequisite for the monofractal interpretation. If pronounced crossovers were present, a single $\alpha$ would represent an oversimplified average of multiple scaling regimes, necessitating multifractal analysis.

\subsection*{Detection Procedure}

\begin{figure}[H]
    \centering
    \begin{minipage}{0.48\textwidth}
        \centering
        \includegraphics[width=\textwidth]{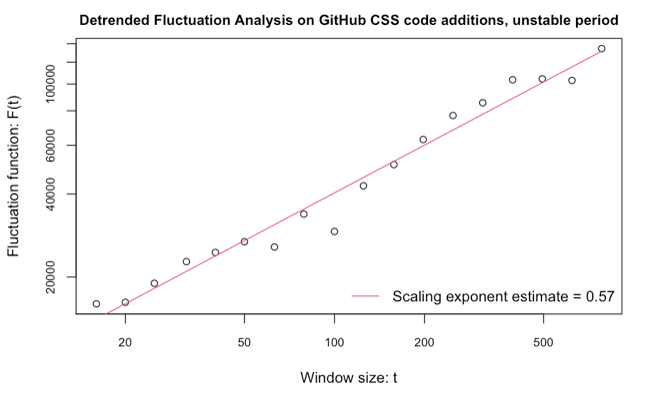}
    \end{minipage}
    \hfill
    \begin{minipage}{0.48\textwidth}
        \centering
        \includegraphics[width=\textwidth]{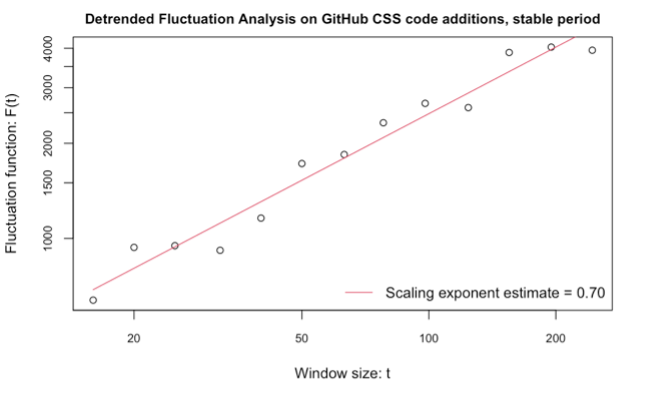}
    \end{minipage}

    \vspace{0.5cm} 

    \begin{minipage}{0.48\textwidth}
        \centering
        \includegraphics[width=\textwidth]{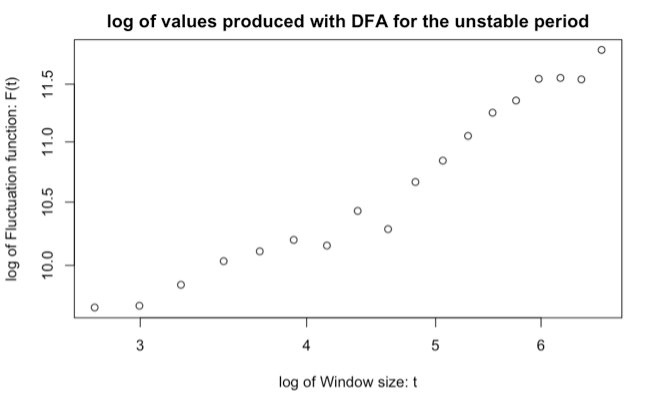}
    \end{minipage}
    \hfill
    \begin{minipage}{0.48\textwidth}
        \centering
        \includegraphics[width=\textwidth]{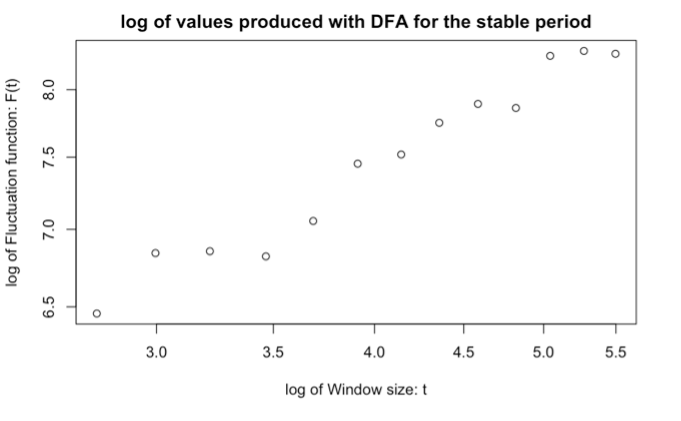}
    \end{minipage}
    
    \caption{DFA log-log plots for stable and unstable periods. The strong overall linearity suggests that the monofractal interpretation is sufficient for comparative analysis.}
    \label{fig:grid_plots}
\end{figure}

\begin{figure}[H]
    \centering
    \includegraphics[width=0.5\textwidth]{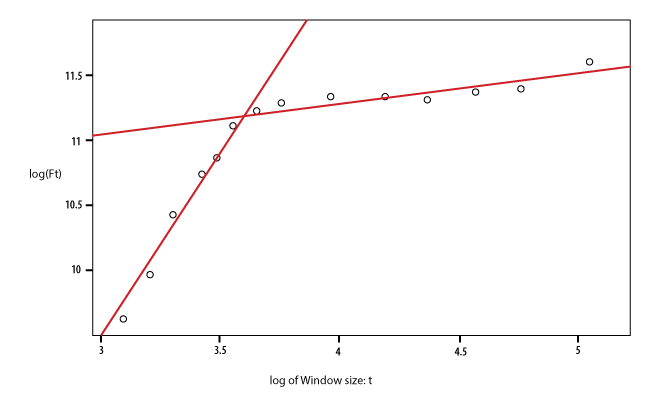}
    \caption{Illustrative example of a definitive crossover, identifying the intersection of two distinct scaling limbs. This diagnostic check confirms that our code-commit data lacks the structural "kink" characteristic of multi-regime systems \citep[adapted from Figure 2 in][]{varela2016}.}
    \label{fig:final_check}
\end{figure}

Crossover detection involves evaluating the linearity of the fluctuation function $F(n)$ on a log-log scale. A robust scaling law is indicated by data points that closely follow a single straight line across the window sizes $n$. The composite results for our datasets are shown in Figure~\ref{fig:grid_plots}, while a representative example of a definitive crossover (for comparison) is shown in Figure~\ref{fig:final_check}.

\subsection*{Results and Validation}

Inspection of the \textsc{dfa} plots (Figure~\ref{fig:grid_plots}) reveals \textbf{no systematic shifts} in the scaling exponent. The scaling behavior is sufficiently consistent for monofractal interpretation, unlike systems where a delayed or blunted crossover signals a fundamental shift in regulatory dynamics \citep{varela2016}.

\end{document}